\documentclass{IEEEtran}
\usepackage{cite}
\usepackage{multirow}
\usepackage{amsmath,amssymb,amsfonts}
\usepackage{graphicx}
\usepackage{textcomp,nicefrac}
\makeatletter
\def\endthebibliography{%
	\def\@noitemerr{\@latex@warning{Empty `thebibliography' environment}}%
	\endlist
}
\makeatother
\def\BibTeX{{\rm B\kern-.05em{\sc i\kern-.025em b}\kern-.08em
T\kern-.1667em\lower.7ex\hbox{E}\kern-.125emX}}
\begin{document}
\title{Evaluation of Xilinx Deep Learning Processing Unit under Neutron Irradiation}
\author{Dimitris Agiakatsikas, Nikos Foutris, Aitzan Sari, Vasileios Vlagkoulis, Ioanna Souvatzoglou, Mihalis Psarakis, Mikel Luján, Maria Kastriotou and Carlo Cazzaniga

\thanks{Experiments  at  the  ISIS  Neutron  and  Muon  Source  were supported  by  a  beamtime  allocation  RB2000230  from  the Science  and  Technology  Facilities  Council.  This  work  has been partially supported by the University of Piraeus Research Center and the EU Horizon 2020 EuroEXA 754337 grant.}
\thanks{Dimitris Agiakatsikas (e-mail: agiakatsikas@gmail.com), Aitzan Sari, Vasileios Vlagkoulis, Ioanna Souvatzoglou, and Mihalis Psarakis (e-mail: mpsarak@unipi.gr) are with the Dept. of Informatics, University of Piraeus, Greece.}
\thanks{Nikos Foutris and Mikel Luján are with the Dept. of Computer Science, The University of Manchester, UK.}
\thanks{Maria Kastriotou and Carlo Cazzaniga are with
the ISIS Facility, STFC, Rutherford Appleton Laboratory, Didcot OX110 QX, UK.}
\thanks{\textcopyright2021 IEEE.  Personal use of this material is permitted.  Permission from IEEE must be obtained for all other uses, in any current or future media, including reprinting/republishing this material for advertising or promotional purposes, creating new collective works, for resale or redistribution to servers or lists, or reuse of any copyrighted component of this work in other works. This paper has been accepted by the 2021 European Conference on Radiation and Its Effects on Components and Systems (RADECS)}
}
\maketitle

\begin{abstract}
This paper studies the dependability of the Xilinx Deep-Learning Processing Unit (DPU) under neutron irradiation. It analyses the impact of Single Event Effects (SEEs) on the accuracy of the DPU running the \texttt{resnet50} model on a Xilinx Ultrascale+ MPSoC.
\end{abstract}


\section{Introduction}
\label{sec:introduction}
\IEEEPARstart{F}{ield} Programmable Gate Arrays (FPGAs) have evolved from glue-logic devices to sophisticated heterogeneous computing platforms that pave the way to the new Artificial Intelligence (AI) wave. 

As a result, datacenter market leaders increasingly integrate FPGA System-on-Chip (SoC) in their infrastructure to target complex AI applications. Companies like Microsoft, Amazon Web Services, and Baidu scale up AI and high-performance applications on hundreds of thousands of Intel and Xilinx FPGA devices \cite{FPGA-cloud-Keller-Wirthlin-2019}. 

A popular application of AI is Convolutional Neural Networks (CNNs).
A growing body of research shows that deploying optimised CNN models on FPGAs achieves a higher performance per watt than CPU and GPU solutions \cite{DNN-approximation-2019}. However, modern FPGAs are vulnerable to Single Event Upsets (SEUs) due to their reliance on SRAM memory to store their configuration and application data. SEUs in SRAM FPGA-SoCs are not destructive but cause various failure modes, such as Silent Data Corruption (SDC), application crashes and kernel panics when an OS is used. 

Previous works have explored the Architectural Vulnerability Factor (AVF) of custom FPGA CNN designs with fault-injection campaigns and irradiation experiments \cite{libano_tns_2019, anderson_redw_2018}. These works have targeted relative simple CNN case studies for edge computing that neither require an Operating System (OS) nor have complex CNN topologies. Simple case studies serve well when focusing on analysing the reliability tradeoffs of various CNN configurations, e.g., the reliability of a CNN under different quantisation and model-compression schemes. Exploring the reliability of large-scale datacenter CNN applications, however, requires a different testing paradigm. The case studies should test additional aspects of a system, such as the operating system's stability, while SEUs are occurring in resources of the CPU, e.g., L1 and L2 caches, on the on-chip Ethernet controller and in custom logic implemented with Programmable Logic (PL).

To this extent, this work analyses the dependability of the whole computational stuck of a commercial CNN inference solution, namely the Xilinx Vitis AI Deep Learning Processing Unit (DPU). In more detail, we implemented the DPU on a Zynq Ultrascale+ XCZU9EG MPSoC, which executed image classification with the \texttt{resnet50} CNN model. By performing Neutron Irradiation experiments, we observed that the FPGA-SoC OS did not crash due to errors in L1 and L2 caches of the CPU, while the DPU application had a very low AVF.  

\section{Background}
\label{sec:background}
\subsection{Vitis AI and the Deep Learning Processing Unit (DPU)}
\label{subsec:DPU}

Xilinx has introduced a rich ecosystem of tools and Intellectual Property (IP) cores to ease the development of AI applications. In more detail, Xilinx provides the Vitis AI development environment that encompasses 1) AI frameworks (e.g., Tensorflow), 2) pre-optimised AI models, 3) quantization and model compression tools, and 4) the  DPU, with all necessary Linux drivers to seamlessly deploy a CNN application on Xilinx devices.

\begin{figure}[t]
\centerline{\includegraphics[width=3.5in]{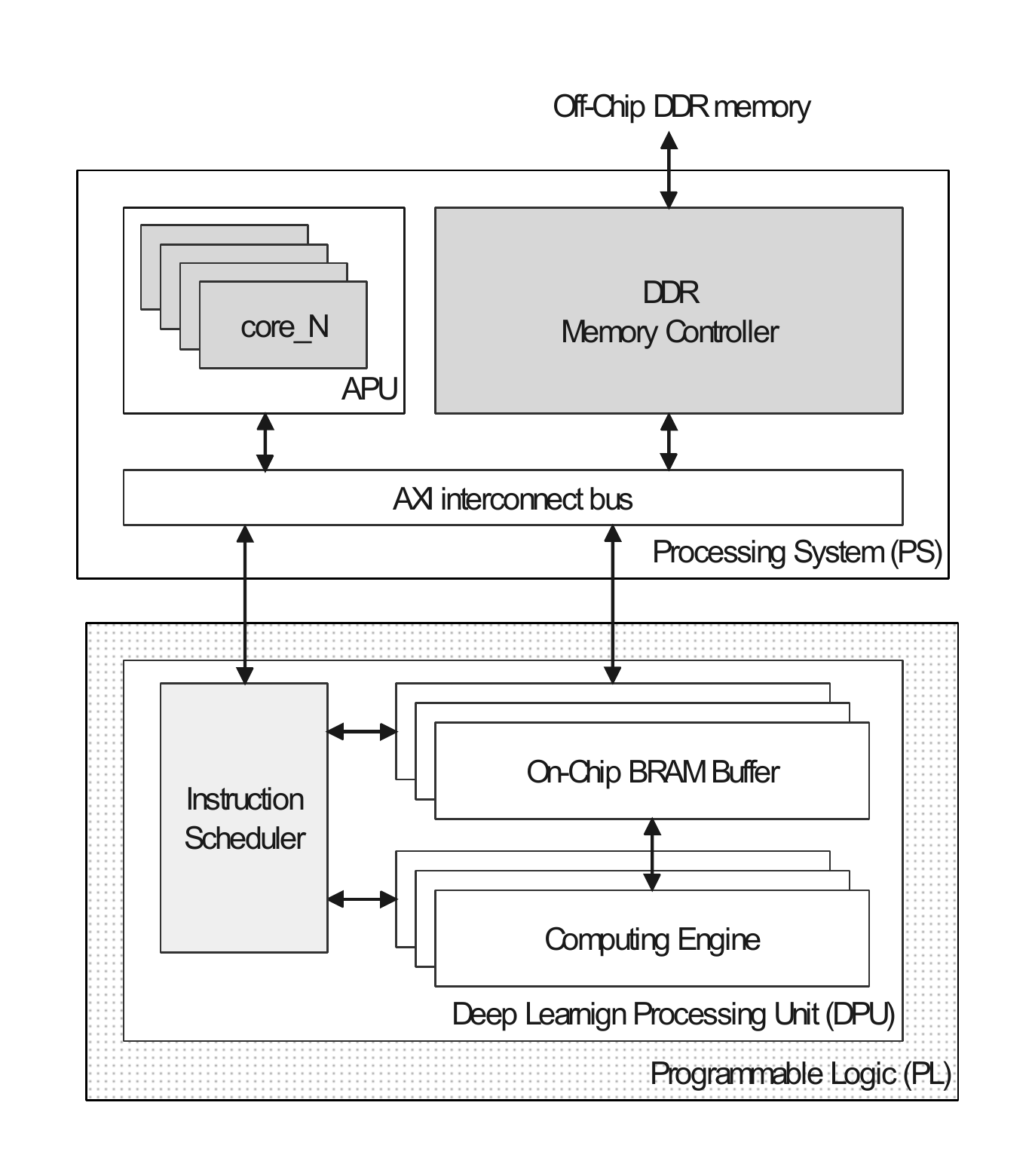}}
\vspace*{-7mm}
\caption{Deep-learing acceleration with the Xilinx Deep Processing Unit (DPU) on Zynq\textsuperscript{\textcopyright}-7000 SoC and Zynq\textsuperscript{\textcopyright} Ultrascale+\textsuperscript{TM} MPSoC devices.}
\label{fig:DPU}
\end{figure}

The DPU is a convolution neural network accelerator IP offered by Xilinx for Zynq-7000 SoC and Zynq Ultrascale+ MPSoC devices. 
The DPU is implemented with PL and is tightly interconnected via the AXI bus to the SoC processing system (PS), as shown in \figurename~\ref{fig:DPU}. The DPU executes special instructions that are generated by the Vitis AI compiler.
A typical Vitis AI development flow involves 1) the optimisation and compilation of a CNN model to DPU instructions and 2) the compilation of software running on the Application Processing Unit (APU), i.e., CPU. 

The APU pre- and post-processes DNN data, controls the DPU, and orchestrates the movement of instructions and data between the DPU, the CPU, and the off-chip DDR memory.

The DPU consists of an instruction scheduler and up to three on-chip BRAM buffers and computing engines. The instruction scheduler fetches and decodes DPU instructions from off-chip memory and controls the on-chip memories and computing engines. The DPU is available in eight configurations, i.e., B512, B800, B1024, B1152, B1600, B2304, B3136, and B4096. Each configuration utilises a different number of computing engines and on-chip memories in order to target different sized devices and support various DPU functionalities, e.g., ReLU, RELU6, or Leaky ReLU.  

\section{Experimental Setup}
\label{sec:Experimental-setup}

\subsection{ChipIr Neutron Beam}
\label{subsec:ChipIR-Neutron-Beam}
ChipIr is an ISIS neutron and muon facility instrument at the Rutherford Appleton Laboratory (UK), designed to deliver an atmospheric-like fast neutron spectrum to test radiation effects on electronic components and devices \cite{chipir1}, \cite{chipir2}. The ISIS accelerator provides a proton beam of \texttt{800~MeV, 40~$\mu$A, 10~Hz}, impinging on the tungsten target of its Target Station 2, where ChipIr is located.
The spallation neutrons produced illuminate a secondary scatterer which optimises the atmospheric-like spectrum arriving at ChipIr, with an acceleration factor of up to \texttt{10\textsuperscript{9}} for ground-level applications. With a frequency of \texttt{10 Hz}, the beam pulses consist of two \texttt{70 ns} wide bunches separated by \texttt{360 ns}. The beam fluence at the position of the Device Under Test (DUT) was continuously monitored by a silicon diode, while the beam flux of neutrons above \texttt{10 MeV} during the experimental campaign was \texttt{5.6~x~10\textsuperscript{6}} neutrons/cm\textsuperscript{2}/s. The beam size was set through the two sets of the ChipIr jaws to \texttt{7cm x 7cm}.

\begin{figure}[t]
\centerline{\includegraphics[width=3.3in]{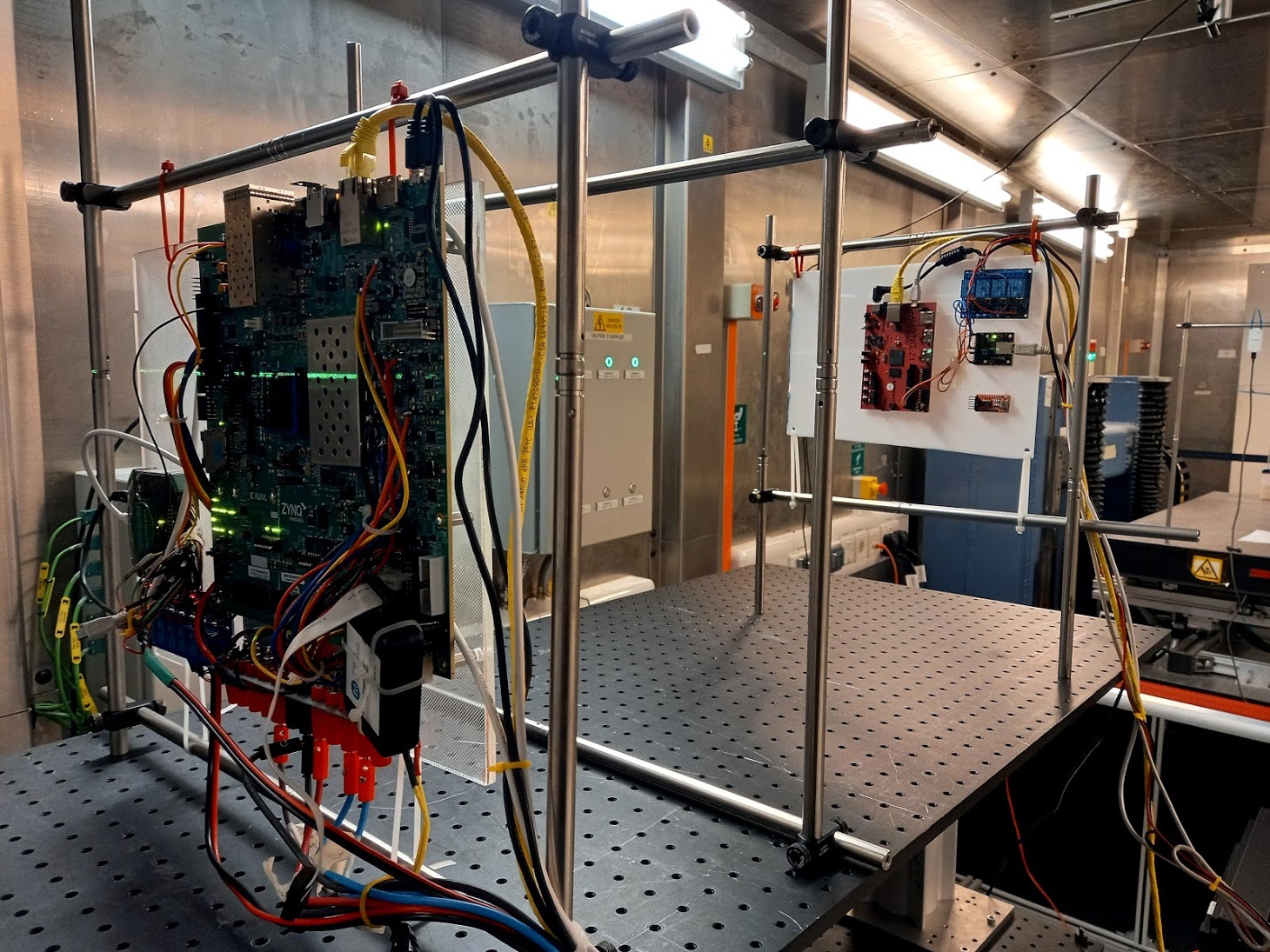}}
\caption{Neutron beam experiment at the ChipIr facility of RAL, UK.}
\label{fig:beam}
\end{figure}

\subsection{Design Under Test (DUT): DPU-B4096}

The Vivado DPU targeted reference design (TRD) \cite{vitis-1.3.1} provided by Vitis AI v1.3.1 was implemented with Vivado 2020.2 for a ZCU102 development board. The ZCU102 board features the Zynq UltraScale+ XCZU9EG MPSoC. The \texttt{B4096} configuration of the DPU was synthesised with default settings, i.e.,  with \texttt{RAM\char`_USAGE\char`_LOW, CHANNEL\char`_AUGMENTATION\char`_ENABLE, DWCV\char`_ENABLE, POOL\char`_AVG\char`_ENABLE, RELU\char`_LEAKYRELU\char`_RELU6, Softmax}.

The design was implemented with Vivado's \texttt{Performance\char`_ExplorePostRoutePhysOpt*} run strategy because Vivado's \texttt{default} run strategy resulted in time violations.
Table~\ref{table:resource-util} shows the resource utilisation and operating frequency of the DPU TRD.
Due to the high resource utilisation of the TRD, Vivado reported a relatively high percentage (\texttt{41.45\%}) of essential bits -- \texttt{59281993} out of the \texttt{143015456} total configuration bits were essential bits. Please recall that essentials bits are configuration bits that, when corrupted have the potential to cause functional errors.
\begin{table}
\centering
\caption{Resource utilisation and operating frequency of the DPU targeted reference design}
\begin{tabular}{|l|r|r|r|r|}
\hline
Resource & \multicolumn{1}{l|}{Utilisation} & \multicolumn{1}{l|}{Available} & \multicolumn{1}{l|}{Utilisation \% } & \multicolumn{1}{p{4.78em}|}{Frequency} \\
\hline \hline
LUT   & 108,208 & 274,080 & 39.48 & 325 MHz \\
LUTRAM & 11,960 & 144,000 & 8.31  & 325 MHz \\
FF    & 203,901 & 548,160 & 37.20 & 325 MHz \\
BRAM  & 522   & 912   & 57.24 & 325 MHz \\
DSP   & 1,395 & 2,520 & 55.36 & 650 MHz \\
IO    & 7     & 328   & 2.13  & 325 MHz \\
BUFG  & 6     & 404   & 1.49  & 325 MHz \\
MMCM  & 1     & 4     & 25.00 & 325 MHz \\
PLL   & 1     & 8     & 12.50 & 325 MHz \\
APU   & 1     & 1     & 100.00 & 1200 MHz \\
DDR ctrl. & 1     & 1     & 100.00 & 533  MHz \\
\hline
\end{tabular}%

\label{table:resource-util}
\end{table}
Two important notes can be made for Table~\ref{table:resource-util}.

1\textsuperscript{st} note: 
All resources in the DPU operate at \texttt{325~MHz} except for the DSPs, which run at \texttt{2~x~325~MHz~=~650~MHz}. This is because the DPU design applies a double data rate technique on DSP resources. Since DSPs are able to operate in a much higher frequency than other PL resources, one can perform \emph{N} times more computation by running the DSPs with \emph{N} times the frequency of the surrounding logic while multiplexing and de-multiplexing their input and output data, respectively. 

2\textsuperscript{nd} note:
The design utilises \texttt{319, 55, 405, 4} and \texttt{1} LUT, LUTRAM, Flip-Flops (FF), BRAM and DSP more resources, respectively, than the baseline \texttt{DPU-TRD} design. This is because we included the Xilinx Soft Error Mitigation (SEM) controller in the design to perform fault injection and validate our experimental setup prior to the radiation tests. The clock of the SEM controller was gated off during beamtime to replicate a simple \emph{out-of-the-box} implementation scenario.

We used Petalinux 2020.2 to generate a Linux OS image for the ZCU102 by using the default Board Support Package (BSP) provided by the \texttt{DPU-TRD}, except 1) the \texttt{nfs\_utils} package which was additionally added in \texttt{RootFS} to mount a Network File Sharing (NFS) folder on Linux, and 2) the u-boot bootloader that mounted an external \texttt{SD EXT4} file system instead of \texttt{INITRD} RAM disk.

The CNN application that run on the DPU was the \texttt{resnet50.xmodel}, which was also provided by the Vitis AI \texttt{DPU-TRD}. This \texttt{resnet50} model is neither compressed nor quantised but serves nice as a baseline application for comparison with more optimised models that we aim to implement and test in future work.

\subsection{Test procedure}

\begin{figure}[t]
	\centerline{\includegraphics[width=3.5in]{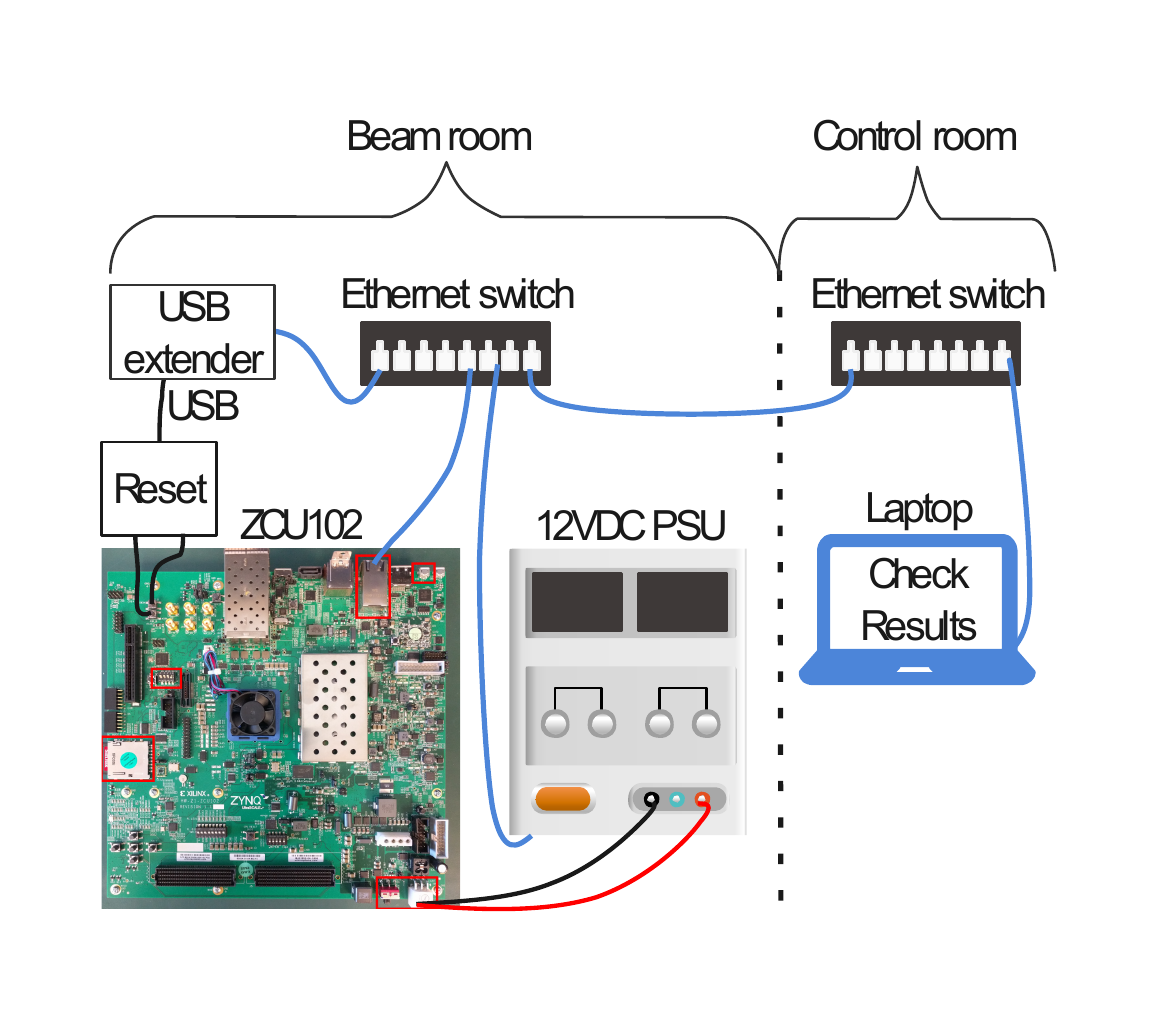}}
	\vspace*{-10mm}
	\caption{Test setup at the ChipIr facility of RAL, UK.}
	\label{fig:setup}
\end{figure}

\figurename~\ref{fig:setup} shows the test setup of the radiation experiment.
A laptop in ChipIr's \textit{control room} orchestrated the test procedure of the DUT. The ZCU102 development board (hosting the DUT), an Ethernet-controlled Power Supply Unit (PSU), and a USB device that remotely reset the ZCU102 (i.e., by electrically shorting the \texttt{SRTS\_B} and \texttt{POR\_B} buttons of the board) was located in the \textit{beam room}. 

The test of the DUT took place as follows.
1) An Experiment Control Software (ECS) running on the laptop remotely resets the DUT and waits for the DUT to boot, 2) the DUT Linux OS restarts, and 3) after a successful kernel boot, an \texttt{/etc/init.d/startup.sh} script executes the following sub-tasks: 4a) the DUT connects on an NFS folder located on the laptop, 4b) the DUT writes a \texttt{sync.log} file in the shared NFS folder to notify the ECS of a successful boot, 4c) an initial \texttt{resnet50} classification takes place to warm-up the CPU caches, 4d) the \texttt{sync.log} is updated to notify ECS that it is ready to start image classifications, 4e) the \texttt{/etc/init.d/startup.sh} enters an infinite loop where it continuously runs DPU classifications and stores the results in the NFS folder to be checked by the ECS. The result checking (i.e. by the ECS) of each classification iteration is synchronised with the DUT via a \textit{mutex} stored in the shared \texttt{sync.log} file. The ECS remotely resets the DUT when it detects a boot timeout, a \textit{Critical Error}  (see Sec.~\ref{sec:Experimental-results}) or a result timeout. It is worth noting that for each classification iteration, the DUT saves the classification result and the Linux \texttt{dmesg.log} for post-analysis. 
\vspace{-2mm}
\vspace{-2mm}
\section{Experimental Results}
\label{sec:Experimental-results}

In this section, we discuss the impact of neutron radiation effects on the reliability of the DPU accelerator. Given that the DPU comprises of a heterogeneous architecture including the ARM SoC and the FPGA fabric, we first present the cross-sections of the memories of the PS part (i.e. CPU caches) and then discuss how the  SEUs in the PL configuration memory affect the behaviour of the system. Please note that the neutron radiation experiments took place at ChipIr on May 2021.

Table~\ref{tab:l1c-cross-section} presents the cross-sections of the 32KB Level-1 Data (L1-D) Cache, the 32KB Level-1 Instruction (L1-I) Cache, and the translation lookaside buffer (TLB) -- a two-level TLB with 512 entries that handles all translation table operations of the CPU. 
Table~\ref{tab:l2c-cross-section} presents the cross-sections of the 1MB Level-2 (L2) Cache and the Snoop Control Unit (SCU). The SCU has duplicate copies of the L1 data-cache tags. It connects the APU cores with the device's accelerator coherency port (ACP) to enable hardware accelerators issue coherent accesses to the L1 memory space.
The upsets in the data and tag arrays in both the L1 and L2 caches have been separately identified. The cross-sections of the tag arrays have been calculated based on the tag sizes of the caches, e.g., a 16-bit tag in the 16-way set associative 1MB L2 cache. 
The cross sections have been calculated for a total fluence of \texttt{5.5x10\textsuperscript{10}} neutrons/cm\textsuperscript{2} on a more than 3-hour radiation experiment.
The results show that the cross-sections of the tag arrays are slightly lower than those of the data arrays. Our cross-section calculations for all caches (i.e., L1 and L2) are very close to those reported in \cite{anderson_redw_2018}. 

\begin{figure}[t]
\centerline{\includegraphics[width=3.5in]{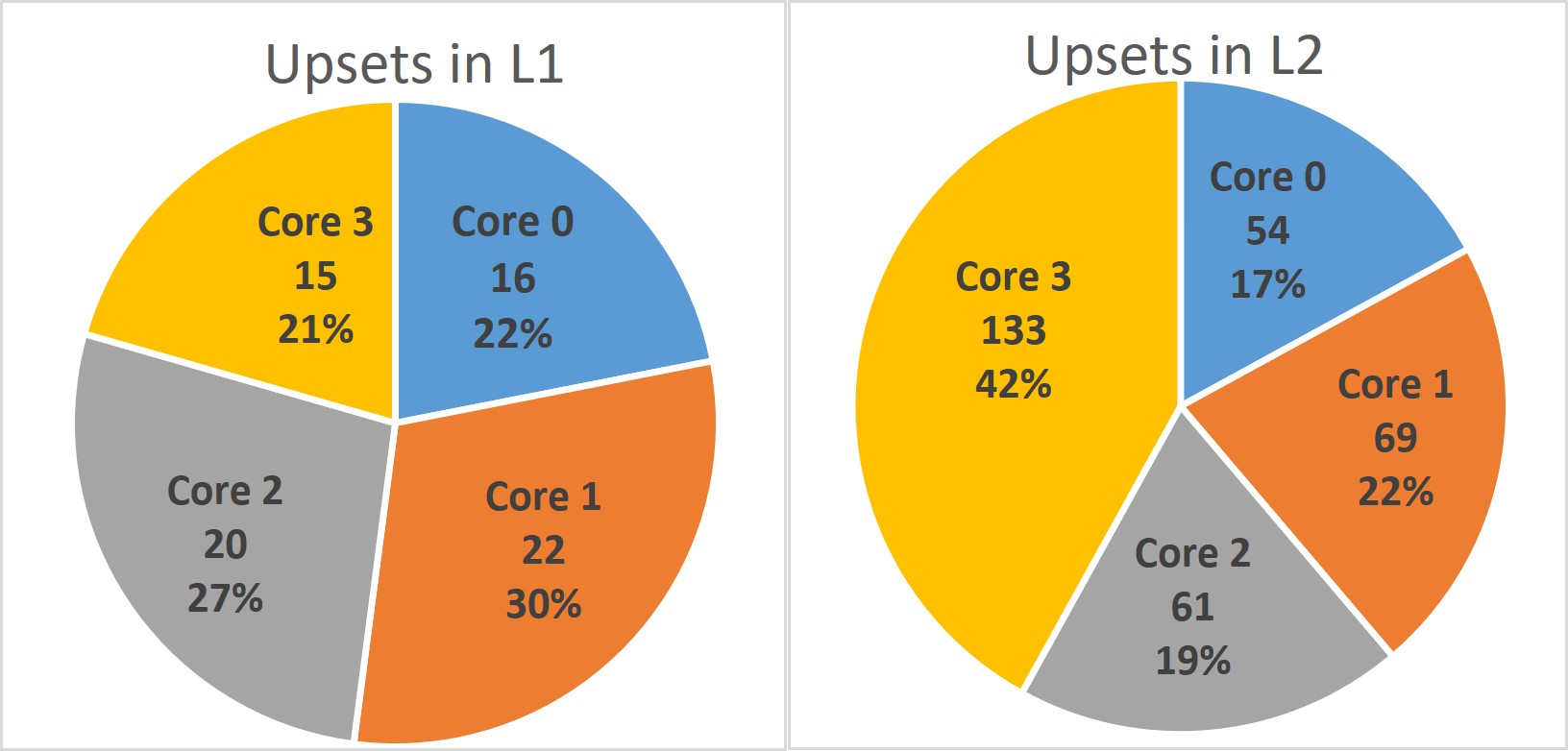}}
\caption{Cache upsets per CPU Core.}
\label{fig:cache-upsets-per-core}
\end{figure}

\figurename~\ref{fig:cache-upsets-per-core} presents the proportion of detected upsets during cache accesses per CPU core. As shown in the figure, the upsets in the L1 caches are balanced between the four cores, while in the L2 cache, more upsets were observed in Core 3. In future work, we aim to save the utilisation of all MPSoC cores and OS running processes to better understand the unbalanced distribution of detected upsets per core in L2.

All MPSoC caches are protected against SEUs with Error Correction Code (ECC) mechanisms, e.g., L1-D and L2 caches incorporate ECC with Single Error Correction Double Error Detection (SECDED) capability, while the L1-I cache has parity that supports only SED. 
All these SED and SECDEC mechanisms in L1 and L2 caches mitigated soft errors in the CPU of the FPGA-SoC, therefore resulting in a stable OS execution, with no application crashes or kernel panics during the radiation tests. This was achieved either by having the ECC scheme correct the error or by flushing and reloading the cache during exceptions of the Linux EDAC driver~(\texttt{/sys/devices/system/edac/mc}).   
\vspace{-2mm}
\begin{table}[htbp]
  \centering
  \caption{L1 Cache Cross Section}
    \begin{tabular}{|l|r|r|r|r|}
          \hline
          & \multicolumn{1}{c|}{Upsets} & \multicolumn{1}{c|}{Cross Section} & \multicolumn{2}{c|}{Conf. Level 95\%} \\
          &  & \multicolumn{1}{c|}{(cm\textsuperscript{2}/bit)} & \multicolumn{1}{c|}{Lower} & \multicolumn{1}{c|}{Upper} \\
          \hline \hline
    L1-D Data & 32    & 2.20E-15 & 1.50E-15 & 3.11E-15 \\ 
    L1-D Tag & 3     & 3.47E-16 & 7.16E-17 & 1.02E-15 \\ 
    L1-D Total & 35    & 1.51E-15 & 1.05E-15 & 2.10E-15 \\ 
    L1-I Data & 25    & 1.72E-15 & 1.11E-15 & 2.54E-15 \\ 
    L1-I Tag & 4     & 4.89E-16 & 1.33E-16 & 1.25E-15 \\ 
    L1-I Total & 29    & 1.28E-15 & 8.54E-16 & 1.83E-15 \\ 
    L1 TLB & 9     & 9.90E-15 & 4.53E-15 & 1.88E-14 \\ \hline
    \end{tabular}%
  \label{tab:l1c-cross-section}%
\end{table}%
\vspace{-2mm}
\begin{table}[htbp]
  \centering
  \caption{L2 Cache Cross Section}
    \begin{tabular}{|l|r|r|r|r|}
          \hline
          & \multicolumn{1}{c|}{Upsets} & \multicolumn{1}{c|}{Cross Section} & \multicolumn{2}{c|}{Conf. Level 95\%} \\
          &  & \multicolumn{1}{c|}{(cm\textsuperscript{2}/bit)} & \multicolumn{1}{c|}{Lower} & \multicolumn{1}{c|}{Upper} \\
          \hline \hline
    L2 Data & 293   & 6.29E-16 & 5.59E-16 & 7.06E-16 \\
    L2 Tag & 20    & 8.59E-17 & 5.25E-17 & 1.33E-16 \\
    L2 Total & 313   & 4.48E-16 & 4.00E-16 & 5.01E-16 \\
    Snoop CU & 4     & 4.63E-16 & 1.26E-16 & 1.19E-15 \\ \hline
    \end{tabular}%
  \label{tab:l2c-cross-section}%
\end{table}%
\vspace{-2mm}
\begin{table}[htbp]
  \centering
  \caption{DPU Cross Section}
    \begin{tabular}{|l|r|r|r|r|r|}
          \hline
          & \multicolumn{2}{c|}{Classification}  & \multicolumn{1}{c|}{Cross} & \multicolumn{2}{c|}{Conf. Level} \\
          & \multicolumn{2}{c|}{runs}  & \multicolumn{1}{c|}{Section} & \multicolumn{2}{c|}{95\%} \\
          & \multicolumn{1}{c|}{\#} & \multicolumn{1}{c|}{\%} 
          & \multicolumn{1}{c|}{(cm\textsuperscript{2})} & \multicolumn{1}{c|}{Lower} & \multicolumn{1}{c|}{Upper} \\
          \hline \hline
    Correct runs & 2964  & 50.27\% & \multicolumn{1}{c|}{-} & \multicolumn{1}{c|}{-} & \multicolumn{1}{c|}{-} \\
    Critical (C)  & 46    & 0.78\% & 8.29E-10 & 6.07E-10 & 1.11E-09 \\
    Tolerable (T) & 2886  & 48.95\% & 5.20E-08 & 5.01E-08 & 5.39E-08 \\
    C+T errors  & 2932  & 49.73\% & 5.28E-08 & 5.09E-08 & 5.48E-08 \\ \hline
    \end{tabular}%
  \label{tab:dpu-cross-section}%
\end{table}%
\vspace{-2mm}

Next, we discuss the impact of radiation effects on the DPU classification accuracy. During the 3-hour experiment, the DPU performed 5896 classification runs. Only a very small portion of runs (0.78\%) resulted in misclassification. Notice that the errors of the \texttt{resnet50} classification results are categorized similarly to \cite{libano_tns_2019} as a) Critical Errors, which lead to misclassification and b) Tolerable Errors, where the errors observed in a result do not affect the classification decision. 
Table~\ref{tab:dpu-cross-section} presents the number of Critical Errors and Tolerable Errors and their cross-sections. 
Based on the SEU vulnerability of the configuration memory and the BRAM reported in \cite{anderson_redw_2018} and the programmable resources of the DPU (i.e. essential configuration bits and BRAMs used), we calculate the rate of upsets affecting the DPU execution. These are 0.14 and 0.55 upsets per classification run in the configuration memory and BRAM contents, respectively. This means that each classification run (which lasts 1.7 seconds), experienced, on average, 0.69 upsets. Notice that since scrubbing was not supported in the experiment, the upsets in the configuration memory were accumulated until the next reset cycle of the system. Moreover, we estimated that more than one upsets were accumulated during the 38 seconds boot and warm-up period of each reset cycle. Thus, for all classification runs, the DPU circuit encountered more than one upsets in the PL memories. As a worst-case analysis, we estimate that the AVF of the DPU accelerator is less than 0.78\%, assuming any single upset in the DPU leads to a critical error.  

\section{Conclusion}
\label{sec:Conclusion}

The neutron radiation experiment demonstrated that the ECC and the interleaving schemes integrated into the UltraScale+ MPSoC caches considerably protect the software stack (OS, pre and post-processing of DNN data, data movement) of the Xilinx Vitis AI DPU from radiation-induced SEUs. It was shown that the most vulnerable part of the DPU is the logic implemented in the FPGA PL. Due to the large amount of utilised PL resources by the DPU, we reasoned that it is likely that the CNN application will be highly vulnerable to SEUs. However, due to the inherent error resiliency of the neural network, only a small portion of SEUs lead to classification errors, resulting in a significantly small AVF. In future work, we aim to perform fault injection experiments in the DPU to obtain a better understanding of its failure mechanisms and propose efficient SEE mitigation approaches to further reduce its AVF. 
\bibliographystyle{IEEEtran}
\bibliography{citations}
\end{document}